\documentclass[aps,physrev,reprint,superscriptaddress]{revtex4-2}

\usepackage{amsmath} 
\usepackage{graphicx}
\usepackage{hyperref}

\begin{document}


\title{Network localization governs social contagion dynamics with macro-level reinforcement}

\author{Leyang Xue}
\affiliation{Department of Systems Science, Faculty of Arts and Sciences, Beijing Normal University
, Zhuhai, 519087, China}
\affiliation{International Academic Center of Complex Systems, Beijing Normal University, Zhuhai, 519087, China}
\affiliation{School of Systems Science, Beijing Normal University, Beijing, 100875, China}

\author{Kai-Cheng Yang}
\affiliation{School of Computing, Binghamton University, Binghamton, NY, USA}
\affiliation{Network Science Institute, Northeastern University, Boston, MA, USA}

\author{Peng-Bi Cui}
\email{cuisir610@gmail.com}
\affiliation{Department of Systems Science, Faculty of Arts and Sciences, Beijing Normal University
, Zhuhai, 519087, China}
\affiliation{International Academic Center of Complex Systems, Beijing Normal University, Zhuhai, 519087, China}
\affiliation{School of Systems Science, Beijing Normal University, Beijing, 100875, China}

\author{Zengru Di}
\affiliation{Department of Systems Science, Faculty of Arts and Sciences, Beijing Normal University
, Zhuhai, 519087, China}
\affiliation{International Academic Center of Complex Systems, Beijing Normal University, Zhuhai, 519087, China}
\affiliation{School of Systems Science, Beijing Normal University, Beijing, 100875, China}



\date{\today}

\begin{abstract}
The spread of ideas, behaviors, and technologies generally depends on feedback mechanisms operating across multiple scales.
Previous studies have extensively examined pairwise transmission and local reinforcement.
However, the role of macro-level social influence---where widespread adoption enhances further adoption---remains understudied.
Here, we focus on a contagion process that incorporates both pairwise interactions and macro-level reinforcement.
We show that the contagion undergoes a shift from continuous to mixed-order transition as macro-level influence exceeds a reinforcement threshold.
Simulations on various real-world networks indicate that network localization governs the contagion outcomes by determining the critical point and the reinforcement threshold.
Building on this insight, we develop a structural metric linking network localization to contagion dynamics, revealing a key trade-off: networks that facilitate weak contagion tend to experience slower diffusion and lower adoption rates, while networks that suppress weak contagions enable faster and more widespread adoption.
These findings challenge the conventional belief that stronger local connectivity uniformly promotes contagion.
\end{abstract}


\maketitle

\section{Introduction}
Contagion processes characterize the dissemination of new ideas, technologies, products, and behaviors, playing a central role in the functioning of socio-economic systems.
Understanding the mechanisms underlying these processes provides key insights into collective decision-making and adoption dynamics across a wide range of disciplines~\cite{Rogers2003,ugander2012structural,iacopini2019simplicial}, including management science, economics, sociology, psychology, and physics~\cite{Chatterjee1990,Berry2018,Gabriel2014}.
Because empirical studies often face substantial confounding factors, modeling approaches have become essential tools for analyzing and predicting social diffusion patterns~\cite{Bass1969,Kiesling2012}.
Network science offers a principled framework to examine contagion pathways and uncover the mechanisms driving emergent collective behaviors in diffusion processes~\cite{cencetti2023distinguishing,jiang2014diffusion}.

Classical contagion models typically assume pairwise transmission with a constant and independent probability between susceptible and adopting individuals~\cite{Pastor2015,bettencourt2006power}.
However, real adoption behaviors frequently require reinforcement from multiple exposures before individuals commit to new behaviors~\cite{centola2007complex,Centola2010, centola2018behavior, Guilbeault2018}.
This has motivated the development of threshold models~\cite{Watts2002}, simplicial and higher-order contagion models~\cite{iacopini2019simplicial,chowdhary2021simplicial}, and hypergraph-based frameworks~\cite{de2020social}.
These models primarily emphasize the role of local reinforcement through interacting neighborhoods~\cite{Cui2018}, communities~\cite{su2018optimal,st2022influential}, simplices, or hyperedges~\cite{iacopini2019simplicial,chowdhary2021simplicial,de2020social,ferraz2023multistability,ferraz2024contagion}, in producing nonlinear adoption patterns.

However, the rise of digital communication, large-scale media exposure, and platform-driven recommendation systems has amplified the influence of macro-level social signals on individual adoption decisions.
Empirical studies have documented abrupt increases in adoption rates driven by visibility, popularity, and network effects at the population scale~\cite{onnela2010spontaneous,toole2012modeling,ma2014consumer,duan2009informational,Jo2023}.
Examples include the rapid spread of online platforms such as Facebook, Twitter, and Threads~\cite{toole2012modeling,Kwak2010,Duarte2024}, as well as viral dissemination of digital content and AI tools~\cite{vosoughi2018spread,Kumar2024}.
These dynamics highlight that the global fraction of adopters can directly reinforce micro-level transmission, thereby creating a positive feedback loop that accelerates adoption.

Macro-level reinforcement can arise from various mechanisms, including psychological conformity to emerging social norms~\cite{Mccoy2014,Young2009,Chen2016,Guilbeault2018}, network effects that enhance product utility in large user bases~\cite{Katz1992,Schoder2000,Rohlfs2003}, and cost reductions driven by economies of scale~\cite{Kiesling2012,Scherer1990,Browning2020}.
Regardless of its origins, such system-level reinforcement couples the global state of adoption back into local transmission dynamics.
Capturing this macro–micro interaction requires a modeling framework that integrates global reinforcement with local contagion processes while accounting for underlying network structure, a connection that remains insufficiently explored in existing contagion models.

To close this gap, we develop a model that captures the interplay between macro-level reinforcement and local pairwise transmission, incorporating varying levels of feedback intensity.
The contagion process is modeled using a Susceptible-Infectious-Recovered (SIR)-like framework~\cite{Pastor2015}, where word-of-mouth effects~\cite{Buttle1998} drive initial adoption.
Macro-level influence is represented as the increasing adoption likelihood as the total number of adopters grows.
Specifically, transmissibility $\beta'$ is defined by two key components: (1) the inherent initial attractiveness of the contagion, $\beta$, and (2) a macro-level social influence or reinforcement term $\alpha \frac{R(t)}{N}$, where $R(t)$ and $N$ denote the number of adopters and the total population size, respectively, and $\alpha$ represents the feedback strength.

Simulations on diverse real-world networks reveal a critical threshold in inherent attractiveness, $\beta_c$, below which outbreaks cannot occur.
When macro-level influence exceeds a critical value $\alpha_c$ (i.e., the reinforcement threshold), the outbreak size at $\beta_c$ exhibits an abrupt jump associated with critical scaling behavior.
This indicates that sufficiently strong global feedback transforms gradual diffusion into an abrupt mixed-order (hybrid) transition.

We identify $\alpha_c$ as the onset of mixed-order criticality and show that it is closely tied to network localization strength, providing a clear structural interpretation.
To quantify this relationship, we propose a localization metric that predicts $\alpha_c$ directly from network topology and reveals a fundamental trade-off: for networks with fixed density, lowering $\beta_c$ inevitably raises $\alpha_c$, and vice versa.
Consequently, strongly localized networks can trigger contagion with low intrinsic attractiveness but restrict the final adoption size, whereas weakly localized networks require higher intrinsic attractiveness yet can sustain a much broader spread.

Our findings challenge classical diffusion theory, which posits that networks with high degree heterogeneity or strong clustering are optimal for lowering contagion thresholds.
Instead, we show that such structural features can localize spreading dynamics and restrict global outbreaks, especially when macro-level reinforcement interacts with network localization.
By bridging micro-level transmission and macro-level reinforcement, our framework provides a unified perspective on how network structure shapes diffusion dynamics, offering new strategies for both promoting beneficial adoption and mitigating undesirable contagion.

\section{Model}
\label{subsec:model}
Consider a socio-economic system represented as a network of $N$ individuals (nodes) connected by $L$ links (edges).
Initially, most nodes are in a susceptible state (denoted by $S$) and can transition to an adopter state (denoted by $A$) when influenced by neighboring adopters.
The adoption occurs with a probability $\beta'$, representing the reinforced transmission probability along local connections.
Additionally, adopters may lose interest and transition to a recovered state (denoted by $R$) with probability $\mu$ per time step.
The contagion dynamics are described as follows:
\begin{eqnarray}\label{eq:eqs1}
	S_i(t-1) + A_j(t-1) & \xrightarrow{\beta'(t)} & A_i(t) +A_j(t),\\
	A_i(t-1) & \xrightarrow{\mu} & R_i(t),
\end{eqnarray}
where $Y_i(t)$ ($Y\in \{S,A,R\}$) denotes the state of node $i$ at time $t$.
Local transmission occurs only between directly connected nodes $i$ and $j$.
The contagion process stops once all adopters have transitioned to the recovered state.

\begin{figure*}[t]
    \centering
    \includegraphics[width=\textwidth]{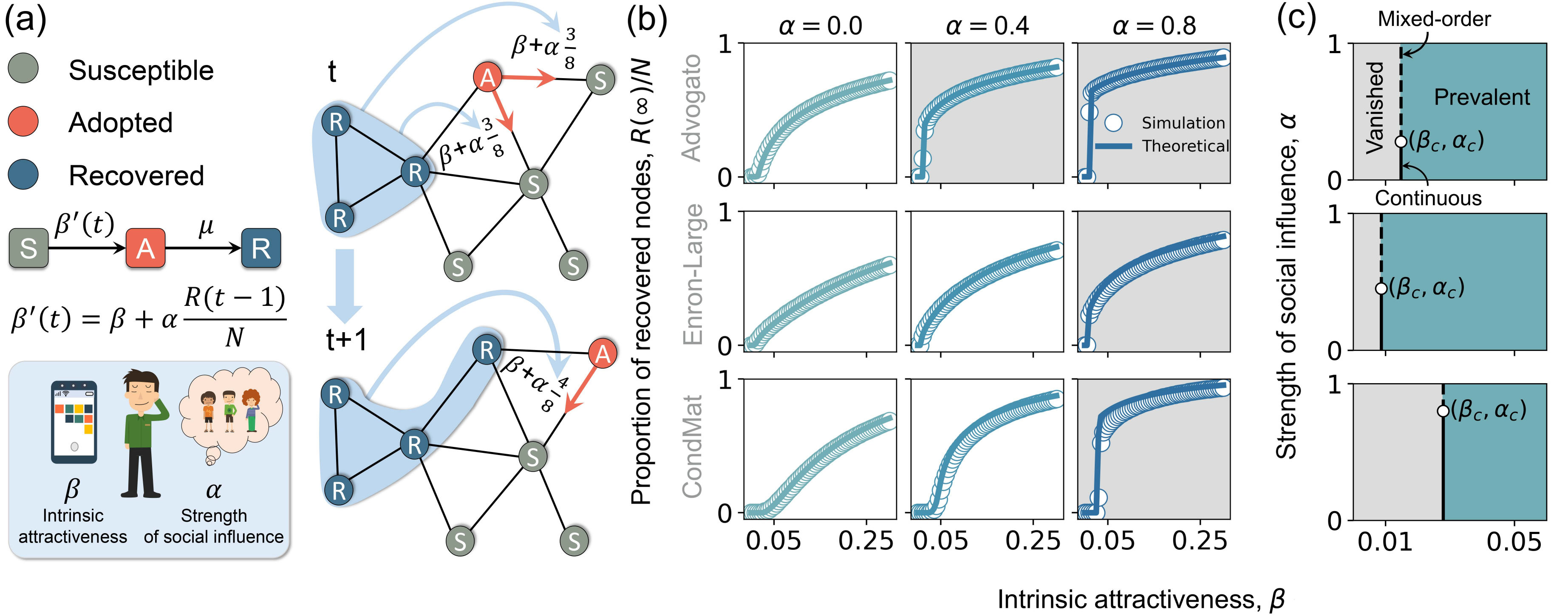}
    \caption{
    \textbf{Definition and behavior of the proposed model}.
    (a) The diagram depicts the contagion process within a simplified network consisting of eight nodes.
    The transmissibility, $\beta'(t)$, of the SIR-like model depends on both intrinsic attractiveness, $\beta$, and macro-level reinforcement, $\alpha\frac{R(t-1)}{N}$.
    At time $t$, an adopter tries to convert two susceptible neighbors, with the macro-level influence contribution to transmissibility being $\alpha \frac{3}{8}$, since there are three recovered nodes in the system.
    At time $t+1$, the level of feedback intensity rises to $\alpha \frac{4}{8}$ due to the addition of one more recovered node.
    (b) The final proportion of recovered nodes, $R(\infty)/N$, is depicted as a function of $\beta$ for three different values of $\alpha$ across three real-world networks (Advogato, Enron-Large, and CondMat networks).
    Dots represent simulation results, while solid lines denote theoretical predictions derived from the dynamic message passing (DMP) method.
    As the level of feedback intensity increases, the contagion exhibits a finite jump in the outbreak size, as indicated by the gray backgrounds.
    Additional details are provided in the main text.
    (c) Phase diagrams for the same networks as in (b) are divided by a vertical line at the critical point $\beta_c$, separating the ``vanished'' and ``prevalent'' contagion states.
    In the former state, the contagion decays, while in the latter state, it spreads widely.
    Solid lines~($\alpha<\alpha_c$) indicate continuous transitions, while dashed curves~($\alpha>\alpha_c$) indicate transitions that develop a finite jump in outbreak size.
    The white dot marks the value of $\alpha_c$ at which this qualitative change in transition behavior occurs.
    }
    \label{fig:1}
\end{figure*}

To incorporate macro-level reinforcement~\cite{Bale2013,Mccoy2014,Delre2010}, we adopt a linear feedback mechanism inspired by the classical Bass diffusion model~\cite{Bass1969}, where the adoption rate increases linearly with the cumulative number of adopters.
This formulation is supported by empirical studies~\cite{Young2009,Bass1969,Rogers2003}, which report approximately linear increases in adoption likelihood during the early and intermediate stages of diffusion.
Although real-world reinforcement mechanisms may display nonlinear effects such as thresholds or saturation, this linear approximation provides a parsimonious yet empirically grounded representation that captures the core features of macro-level reinforcement while maintaining analytical tractability for exploring the role of network structure in contagion dynamics.

Specifically, the time-dependent transmissibility $\beta'$ is defined as:
\begin{eqnarray}
	\label{eq:eqs3}
	\beta'(t) =  \min \left(1,\;\beta + \alpha \frac{R(t-1)}{N}\right), \alpha \geq 0, \beta' \in \left[ 0,1\right].
\end{eqnarray}
Here, $\beta$ represents the intrinsic attractiveness of the contagion~\cite{Rogers2003}, while $\alpha\frac{R(t-1)}{N}$ captures the macro-level influence from cumulative adoption.
As $R(t-1)$ increases, the probability that a susceptible individual adopts also increases, reflecting the reinforcing effect of collective behavior.
The parameter $\alpha$ controls the level of feedback intensity, and the upper bound ensures that $\beta'(t)$ remains within the valid probability range.
An overview of the model is illustrated in Fig.~\ref{fig:1}(a).

\section{Result}
\subsection{Phase Transition}
To characterize the behavior of our model, we perform simulations across a range of $\beta$ and $\alpha$ values on various real-world networks.
Results for three representative networks are shown in Fig.~\ref{fig:1}(b).
Additionally, we apply the dynamic message passing (DMP) method to theoretically track the final market penetration, i.e., $\frac{R(\infty)}{N}$.
The theoretical predictions exhibit strong agreement with simulation results.
The simulation procedures and the DMP method are described in Appendices A and C, respectively, while information on the network datasets and additional results is provided in the Supplemental Material~\cite{SM1}.

Through extensive analysis, we identify key phenomena that shed light on the contagion process.
In the absence of macro-level reinforcement (i.e., $\alpha=0$), our model reduces to the classical SIR model.
As $\beta$ surpasses a critical threshold $\beta_c$, the system undergoes a continuous transition from the ``vanished'' state, where contagion fails to spread, to the ``prevalent'' state, where a finite fraction of individuals eventually adopt.
The introduction of macro-level reinforcement fundamentally alters these dynamics.
When this feedback strength is sufficiently high, the final adoption size at $\beta_c$ exhibits an abrupt jump, while still displaying critical scaling behavior near the critical point (see Supplemental Material~\cite{SM2}, Sec.~B.2).

This qualitative change reflects the emergence of a mixed-order transition, in which an abrupt jump coexists with critical behavior rather than a purely first-order transition.
In particular, the intrinsic threshold $\beta_c$ remains unchanged across different values of $\alpha$ (see Supplemental Material~\cite{SM3}, Sec.~B.3), since macro-level reinforcement only becomes operative once adopters already exist, requiring $\beta \ge \beta_c$ for contagion to spread initially.

We further identify a reinforcement threshold, $\alpha_c$, beyond which contagion exhibits a mixed-order phase transition (see plots with gray backgrounds in Fig.~\ref{fig:1}(b))~\cite{Boccaletti2016,D2019}.
We refer to this threshold as the onset of mixed-order criticality and provide a numerical estimation method in Appendix D.
The model's behavior is thus characterized by the tuple $(\beta_c, \alpha_c)$, which partitions the phase space into distinct regions, as shown in Fig.~\ref{fig:1}(c).
These thresholds have important real-world implications: $\beta_c$ represents the intrinsic attractiveness required to trigger an outbreak, while $\alpha_c$ determines the level of feedback intensity needed to induce an abrupt but hybrid global adoption surge.
Networks with lower $\alpha_c$ values generally exhibit faster diffusion
and higher adoption rates (see Supplemental Material~\cite{SM4}, Sec.~B.1).

\begin{figure*}[!ht]
    \centering
    \includegraphics[width=\textwidth]{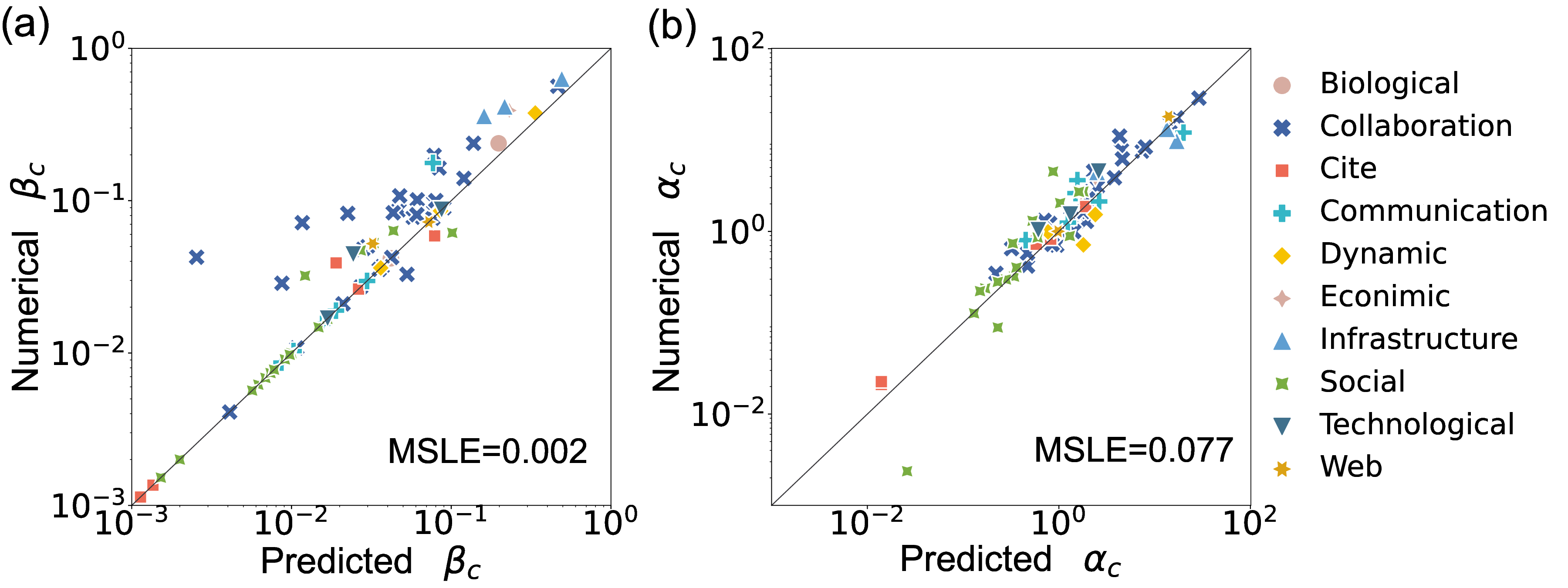}
    \caption{
    \textbf{Critical point~($\beta_c$) and reinforcement threshold~($\alpha_c$) for real-world networks}.
    (a) Comparison between simulated and predicted values of $\beta_c$, based on Eq.~(\ref{eq:eqs23}).
    (b) Comparison between simulated and predicted values of $\alpha_c$, derived from Eq.~(\ref{eq:alpha_l}).
    Each dot represents a real-world network, and the solid diagonal lines ($y=x$) indicate where predicted values match simulated results.
    The mean squared logarithmic errors (MSLE) between simulations and predictions are annotated in each plot.
    }
    \label{fig:2}
\end{figure*}

\subsection{Quantifying the critical point and reinforcement threshold of phase transition}
Given the critical roles of $\beta_c$ (the critical point) and $\alpha_c$ (the reinforcement threshold), accurately estimating their values is essential to understand contagion dynamics in networked populations.
While these thresholds can be determined numerically, the process requires extensive simulations, making it prohibitively expensive (see Appendix A for details).
Here, we demonstrate that both $\beta_c$ and $\alpha_c$ are inherently determined by the network structure and can be efficiently calculated using simple algebraic expressions derived from the non-backtracking matrix $\bm{B}$~\cite{Travis2014}.

In the previous section, we established that the simulated value of $\beta_c$ is independent of the macro-level feedback.
Therefore, it can be approximated by the epidemic threshold of the SIR model~\cite{Koher2019}:
\begin{eqnarray}\label{eq:eqs23}
	\beta_c = \frac{\mu}{\rho(\bm{B})+\mu-1},
\end{eqnarray}
where $\rho(\bm{B})$ is the largest eigenvalue of the non-backtracking matrix $\bm{B}$.
For simplicity, we set $\mu=1$, yielding a threshold approximation of $1/\rho(\bm{B})$.
To validate this approximation, we compare the predicted and simulated values of $\beta_c$ across 74 real-world networks, as shown in Fig.~\ref{fig:2}(a).
The results demonstrate a strong agreement, with minor deviations consistent with previous reports~\cite{Pastor2020}.

The case for $\alpha_c$ is more complex.
When macro-level reinforcement is sufficiently strong, the final adoption size at $\beta_c$ exhibit a finite jump, reflecting an avalanche-like activation of susceptible individuals once $\beta$ surpasses the threshold.
This abrupt collective response arises when a positive feedback loop forms between the increasing number of adopters and the reinforced transmissibility $\beta'$.
However, the presence of macro-level reinforcement alone is insufficient to produce such behavior.
For an avalanche to occur, a substantial fraction of individuals must have similar adoption probabilities, enabling many of them to become activated nearly simultaneously.
These probabilities vary across individuals due to differences in network connectivity~\cite{Pastor2015,Cai2015,Cui2018}.
For example, well-connected individuals are more likely to be exposed to contagion and adopt compared to those with fewer connections.
Previous studies suggest that adoption probability can be approximated using the non-backtracking centrality~\cite{Shrestha2015,Timar2022}, defined as:
\begin{eqnarray}\label{eq:nbc}
    x_i = \sum_j A_{ij} v_{j\rightarrow i},
\end{eqnarray}
where $x_i$ represents the centrality of node $i$, $A_{ij}$ is an element of the adjacency matrix $\bm{A}$, and $v_{j\rightarrow i}$ denotes the component of the principal eigenvector associated with the non-backtracking matrix $\bm{B}$.

\begin{figure*}
  \centering
    \includegraphics[width=\textwidth]{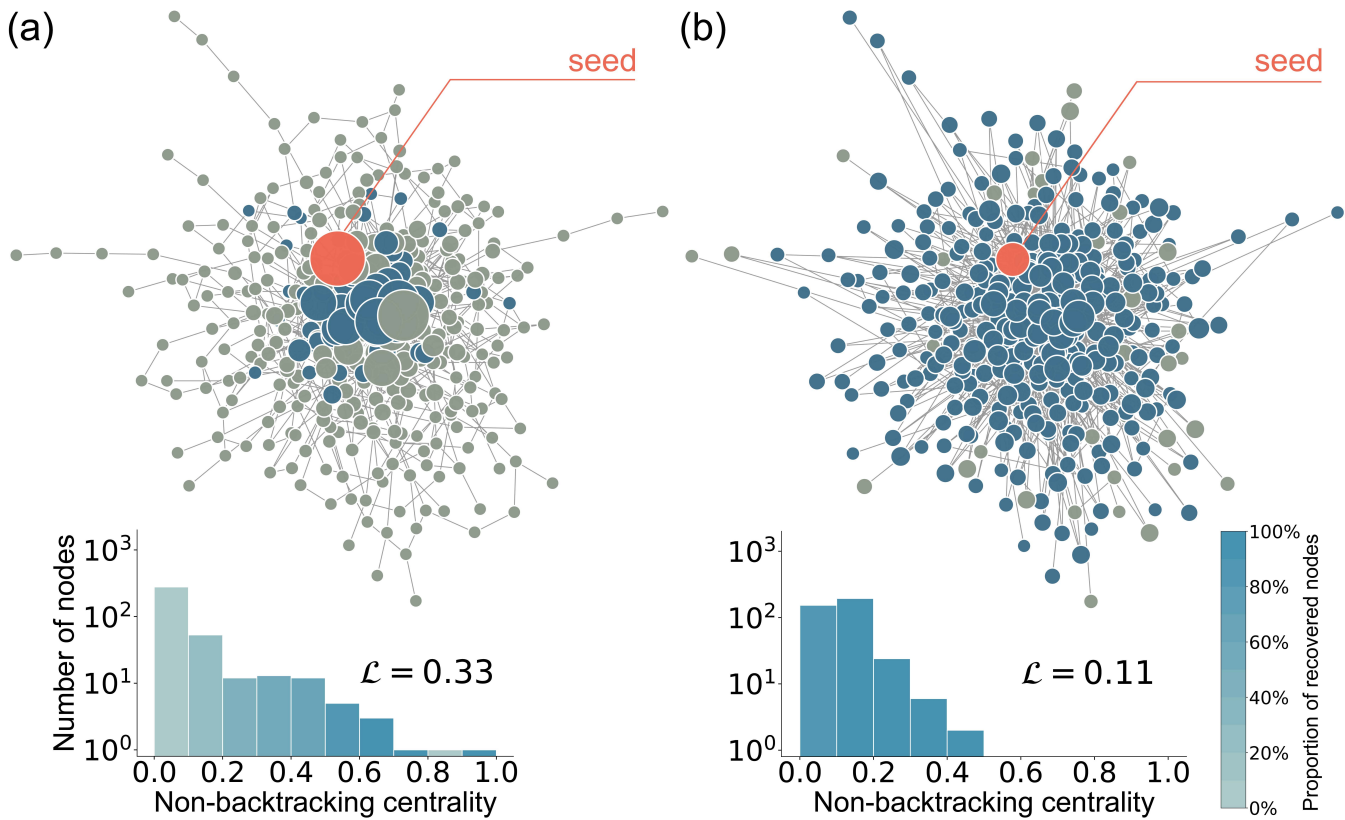}
    \caption{
    \textbf{Effect of network localization on the contagion process}.
    We simulate the proposed model with the parameters $\alpha=1$ and $\beta=\beta_c$ (the theoretical critical value) on two synthetic networks, shown in (a) and (b), respectively.
    The network in (b) is generated by randomly rewiring the edges of the real-world network in (a), while preserving its degree sequence.
    Node sizes correspond to non-backtracking centrality, with gray nodes indicating the final susceptible states and blue nodes indicating the final recovered states; the initial seed is marked in orange.
    The histograms display the distribution of node non-backtracking centrality indices, with each bar's color reflecting the proportion of recovered nodes within the respective bin.
    The figure also includes annotations for the localization metric, $\mathcal{L}$, for both networks.
    For the critical thresholds, we observe $\beta_c=0.104$, and $\alpha_c=1.17$ for network (a), and $\beta_c =0.183, \alpha_c = 0.58$ for network (b).
    }
    \label{fig:3}
\end{figure*}

Based on this heuristic, mixed-order transition behavior is more likely to occur in networks with smaller variances in their node centrality measures.
To illustrate this, we present diffusion outcomes on two synthetic networks in Fig.~\ref{fig:3}.
While these networks share the same degree sequence, differences in their connectivity patterns result in distinct non-backtracking centrality distributions, the reinforcement threshold $\alpha_c$, and final states.
In particular, the network in Fig.~\ref{fig:3}(a) has a small subset of highly central nodes, while the rewired network in Fig.~\ref{fig:3}(b) displays a more homogeneous centrality distribution.
By definition, network (a) has stronger localization strength than network (b).
Consequently, in network (a), recovered nodes tend to be more central, with diffusion concentrated around highly connected nodes, whereas in network (b), contagion spreads more evenly across the network.

To quantify this property, we define the localization strength $\mathcal{L}$ as:
\begin{eqnarray}\label{eq:localization_strength}
    \mathcal{L} = \frac{\sqrt{\frac{1}{N}\sum_{i=1}^{N}(x_i-\langle x_i \rangle)^2}}{\langle x_i \rangle \langle k \rangle},
\end{eqnarray}
where the numerator represents the standard deviation of node non-backtracking centrality measures.
The denominator, given by the product of the average centrality and the average degree, ensures comparability of $\mathcal{L}$ across different networks.
The value of $\mathcal{L}$ for both networks in Fig.~\ref{fig:3} illustrates how this metric quantifies localization strength.

By analyzing the values of $\mathcal{L}$ and $\alpha_c$ across different networks, we observe a positive correlation between them, which we capture with the empirical relationship:
\begin{eqnarray}\label{eq:alpha_l}
   \alpha_c = \lambda \mathcal{L}^\eta,
\end{eqnarray}
where $\lambda$ and $\eta$ are parameters determined empirically.
To avoid overfitting on real-world networks, we fit Eq.~(\ref{eq:alpha_l}) using a collection of synthetic networks, resulting in $\lambda = 2.63$ and $\eta = 1.00$ (see Appendix D for details).
We then validate this relationship by comparing the simulated $\alpha_c$ values with their predicted counterparts, $2.63\mathcal{L}$, across our collection of real-world networks in Fig.~\ref{fig:2}(b), finding excellent agreement.
This relationship enables prediction of $\alpha_c$ for any network based solely on its topology, eliminating the need for extensive simulations.
For practical applications, Eq.~(\ref{eq:alpha_l}) can also be fitted directly on specific real-world networks to refine the estimates of $\lambda$ and $\eta$.

We acknowledge that alternative methods exist for quantifying network localization strength.
For instance, the inverse participation ratio (IPR) is widely used in the literature for this purpose~\cite{Travis2014}.
Other measures include substituting $x_i$ in Eq.~(\ref{eq:localization_strength}) with eigenvector centrality~\cite{Bonacich1972}, or using the Gini coefficient to measure the variance~\cite{Bendel1989}.
However, our definition in Eq.~(\ref{eq:localization_strength}) provides
the most accurate predictions for $\alpha_c$ (see Supplemental
Material~\cite{SM5}, Sec.~B.4 for a comparison with alternative
definitions).

\begin{figure*}
    \centering
    \includegraphics[width=0.6\textwidth]{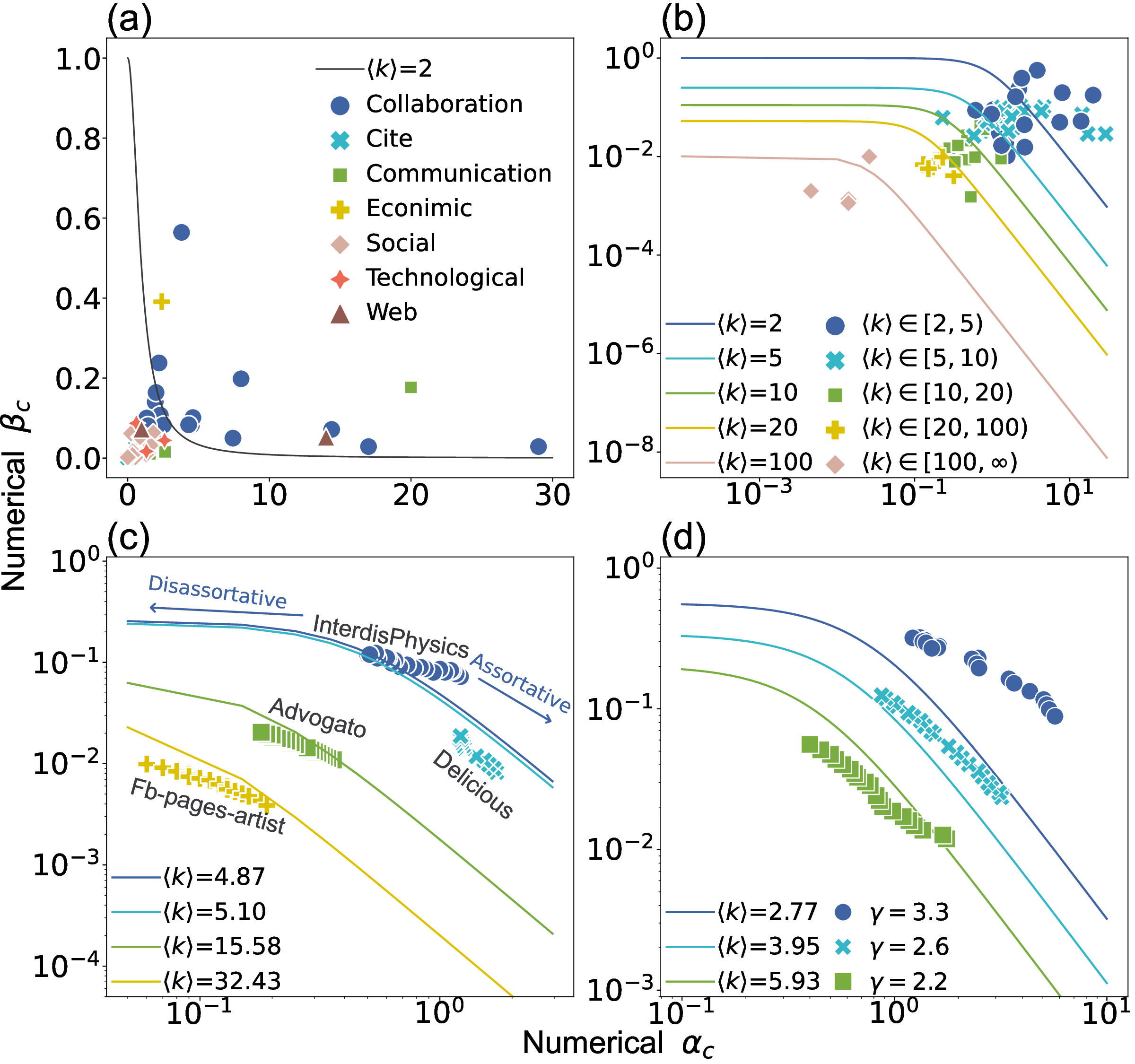}
    \caption{
    \textbf{Relationship between the critical point and the reinforcement threshold}.
    Values of $\beta_{c}$ and $\alpha_{c}$ are obtained through numerical simulations.
    Each marker represents a network, with its position indicating the corresponding intrinsic threshold and reinforcement threshold.
    Solid lines illustrate the relationship between $\beta_c$ and $\alpha_c$ as described by Eq.~(\ref{eq:beta_alpha_relation}) for the specified average degree.
    (a) Equation~(\ref{eq:beta_alpha_relation}) is plotted with $\langle k \rangle = 2$.
    (b) Networks are categorized by average degree, with lines representing different $\langle k \rangle$ values.
    A logarithmic scale is used to highlight finer details.
    (c) Adjusting network assortativity yields configurations with varying $\beta_c$ and $\alpha_c$ values for four selected real-world networks (InterdisPhysics, Delicious, Advogato, Fb-pages-artist).
    Arrows indicate the directions of these changes.
    (d) Similar to (c), but applied to three scale-free networks generated by the configuration model.
    }
    \label{fig:4}
\end{figure*}

Beyond predicting the reinforcement threshold, Eq.~(\ref{eq:alpha_l}) allows us to analytically examine the relationship between $\beta_c$ and $\alpha_c$.
For random uncorrelated networks, we have the following relationship:
\begin{eqnarray}\label{eq:beta_alpha_relation}
   \beta_c = \frac{1}{\langle k\rangle^3 (\alpha_c/\lambda)^{2/\eta} +\langle k \rangle -1},
\end{eqnarray}
where the average degree $\langle k\rangle$ quantifies the network's edge density (see Appendix E for the derivation of Eq.~(\ref{eq:beta_alpha_relation})).
Since all networks considered satisfy $\langle k\rangle > 2$, Eq.~(\ref{eq:beta_alpha_relation}) indicates that $\beta_c$ decreases monotonically with $\alpha_c$, and that higher $\langle k \rangle$ values lead to lower values of both $\beta_c$ and $\alpha_c$.

As previously noted, a smaller $\beta_c$ allows contagion with lower attractiveness to spread, whereas a smaller $\alpha_c$ enables faster diffusion and higher adoption.
Thus, networks with lower $\beta_c$ and $\alpha_c$ values are generally more effective at facilitating broad diffusion.
To assess real-world networks' efficacy, we plot them in the ($\beta_c$, $\alpha_c$) plane in Fig.~\ref{fig:4}(a), including a reference curve from Eq.~(\ref{eq:beta_alpha_relation}) with $\langle k \rangle=2$, which serves as an upper bound.
Most networks align closely with or fall below this curve, supporting our theoretical framework.
Many networks cluster near the origin, suggesting that real-world networks tend to be optimized for social contagion.
There are a few exceptions, primarily among collaboration networks, which may be hindered by cultural or knowledge barriers in knowledge transfer~\cite{Sheng2013}.

To better visualize these relationships, Fig.~\ref{fig:4}(b) reproduces Fig.~\ref{fig:4}(a) using a logarithmic scale, incorporating troughs from Eq.~(\ref{eq:beta_alpha_relation}) with varying $\langle k \rangle$ values and color-coded markers representing network average degrees.
Most networks fall within the predicted troughs, validating our theory on the relationship between the intrinsic threshold $\beta_c$ and the reinforcement threshold $\alpha_c$.
As shown earlier, $\beta_c$ and $\alpha_c$ regulate the contagion processes: to maximize the number of adopters, one would need to minimize both of them.
However, Eq.~(\ref{eq:beta_alpha_relation}) suggests that this is only achievable by increasing the density of the networks (i.e., adding new connections).
When the network density is fixed, Eq.~(\ref{eq:beta_alpha_relation}) introduces a trade-off: increasing localization strength (larger $\mathcal{L}$) would reduce $\beta_c$ but increase $\alpha_c$, and vice versa.

Under this constraint, we show that rewiring the network while preserving edge density can adjust the values of $\beta_c$ and $\alpha_c$.
A particularly effective method is the Xulvi-Brunet–Sokolov algorithm~\cite{Xulvi2004},
which modifies the localization strength by tuning network assortativity
(see Supplemental Material~\cite{SM7}, Sec.~B.7).
Increasing assortativity enhances localization strength, raising $\alpha_c$ while lowering $\beta_c$.
Figure~\ref{fig:4}(c) illustrates this effect on four networks, confirming that Eq.~(\ref{eq:beta_alpha_relation}) holds.
This approach extends to synthetic networks, as shown in Fig.~\ref{fig:4}(d).
Additionally, scatter plots of the predicted $\alpha_c$ and $\beta_c$ values
exhibit patterns consistent with Fig.~\ref{fig:4}, further validating our
theoretical framework regarding the relationship between them (see
Supplemental Material~\cite{SM6}, Sec.~B.6).

\section{Discussion}
In this paper, we present a comprehensive analysis of a contagion process that explicitly couples micro-level pairwise transmission with macro-level positive feedback.
Without macro-level reinforcement, the system undergoes a continuous transition from a vanished state to a prevalent state as the intrinsic attractiveness of the contagion increases.
Once the reinforcement strength exceeds a threshold $\alpha_c$, the final outbreak size exhibits an abrupt jump that is nevertheless accompanied by critical behavior, a hallmark of mixed-order criticality.
This demonstrates the profound impact of macro-level reinforcement on contagion dynamics.

To contextualize these findings within the broader landscape of contagion research, we note that abrupt or explosive epidemic growth has been observed in other irreversible spreading models, including those driven by higher-order interactions~\cite{Malizia2025} and collapse phenomena in limited-resource detection processes~\cite{LamataOtin2023}.
In these systems, strongly nonlinear outbreak curves emerge even though the underlying phase transition remains continuous.
Our work complements these studies by revealing a distinct mechanism that produces mixed-order transitions---characterized by an abrupt order-parameter jump accompanied by critical behavior---through the interplay between global reinforcement and structural localization.

Based on this insight, we identify network localization strength as a key structural factor governing the onset of mixed-order behavior.
To quantify this property, we introduce the metric $\mathcal{L}$, derived from non-backtracking centrality, which captures heterogeneity in node-level infection probabilities for percolation-like (irreversible) dynamics.
By accounting for path redundancy and avoiding overcounting from short cycles, $\mathcal{L}$ enables accurate prediction of the reinforcement threshold $\alpha_c$ directly from network topology.
Its predictive power stands in contrast to classical localization measures such as the inverse participation ratio (IPR)~\cite{Goltsev2012}, which were designed for linear SIS processes and are less suited to percolation-like contagion with feedback.
Across diverse networks, $\mathcal{L}$ consistently outperforms IPR and other localization metrics in estimating $\alpha_c$.
Its algebraic simplicity also suggests its potential applicability to other dynamical processes on networks.

Beyond predicting the onset of mixed-order behavior, $\mathcal{L}$ also reveals a quantitative relationship between the intrinsic critical point $\beta_c$ and the reinforcement threshold $\alpha_c$.
This relationship provides a practical framework for evaluating and optimizing contagion efficiency in socio-economic systems.
When promoting contagion, low values of both thresholds are desirable and can be facilitated by increasing network edge density.
When edge density is fixed, however, a fundamental trade-off emerges: lowering one threshold inevitably raises the other.
Consequently, highly attractive contagions benefit from targeting populations with large $\beta_c$ (and thus smaller $\alpha_c$),  while less attractive contagions can still achieve global spread in populations with lower $\beta_c$, albeit more gradually.

Finally, although our analysis focuses on linear macro-level reinforcement, real contagion processes often involve nonlinear forms of global influence, including thresholds, saturation, and diminishing returns.
Linear reinforcement nevertheless serves as a standard first-order approximation that maintains analytical tractability while exposing the fundamental interaction between feedback and topology.
Therefore, this framework establishes a solid foundation for future extensions incorporating nonlinear reinforcement functions in networked diffusion systems.

\begin{acknowledgments}
This work was supported by the Key Program of the National Natural Science Foundation of China (Grant No.~71731002), and by the Guangdong Basic and Applied Basic Research Foundation (Grant Nos.~2024A1515012692).
Leyang Xue acknowledges the support of the China Scholarship Council Program and the Israeli Sandwich Scholarship.
The authors thank Claudio Castellano for the helpful comments and suggestions.
\end{acknowledgments}

\appendix
\section{Numerical simulations}
In each realization, one node is randomly assigned the $A$ (adopter) state to initiate the diffusion process, while all other nodes are set to the $S$ (susceptible) state.
The diffusion then proceeds according to the model rules until it reaches an absorbing state, where no further transmission is possible.
Throughout this study, the recovery probability $\mu$ is fixed at $1$ for simplicity.

In our model, the critical point $\beta_c$ is determined numerically by identifying the maximum value of susceptibility, defined as
\begin{equation}\label{eq:chi}
\chi =\frac{\sqrt{\langle r(\infty)^2\rangle  - \langle r(\infty) \rangle^2}}{\langle r(\infty) \rangle}.
\end{equation}
To ensure reliability, we perform $10^4$ independent realizations to compute the first and second moments of $r(\infty)$ for a given network under identical parameters.

To determine $\alpha_c$, we classify phase transitions by examining the mass distribution of $r(\infty)$ at the predicted $\beta_c$ values, varying $\alpha$ across numerous independent realizations.
Abrupt transitions are identified by the appearance of an isolated peak
representing giant recovered clusters~\cite{Cai2015}. 
Additional details and illustrative examples are provided in the Supplemental
Material~\cite{SM51}, Sec.~B.5.

\section{Non-backtracking matrix and non-backtracking centrality}
Given the central role of the non-backtracking matrix in our theoretical analysis, we provide a brief introduction here.
In an undirected network with $L$ edges, the non-backtracking matrix $\bm{B}$ is a $2L \times 2L$ non-symmetric matrix in which rows and columns correspond to directed edges $j \rightarrow i$, representing edges directed from node $j$ to node $i$.
The elements of $\bm{B}$ are defined as follows:
\begin{eqnarray}\label{eq:eqs19}
	B_{z \rightarrow i,j\rightarrow z'} = \left \{
	\begin{aligned}
		1   \qquad    &  if \quad z' = z, j \neq i,\\
		0   \qquad     & otherwise. \\
	\end{aligned}
	\right.
\end{eqnarray}

When all nodes in the network belong to a strongly connected giant component, the non-backtracking matrix is non-negative and irreducible.
According to the Perron-Frobenius theorem~\cite{Horn2012}, there exists a positive leading eigenvector $v_{j \rightarrow i}$ associated with the largest eigenvalue $\lambda_B$:
\begin{equation}
    \lambda_B v_{j \rightarrow i} = \sum_{k\rightarrow m} B_{j\rightarrow i,k\rightarrow m} v_{k\rightarrow m}.
\end{equation}
The element $v_{j \rightarrow i}$ of this leading eigenvector represents the centrality of node $j$, excluding any influence from node $i$.
The non-backtracking centrality of node $i$ is then defined as:
\begin{equation}\label{eq:eqs25}
    x_i = \sum_{j}A_{ij}v_{j \rightarrow i}.
\end{equation}
Unlike eigenvector centrality, non-backtracking centrality mitigates the ``self-inflating'' effect associated with hub nodes~\cite{Travis2014,Pastor2020}.

\section{Dynamic Message Passing method}
\label{subsubsec:dmp}
To analytically determine the outbreak size and threshold, we use the dynamic message passing (DMP) method.
DMP is widely applied in contagion studies as it prevents backtracking to the source node, thereby avoiding mutual transmission effects~\cite{Karrer2010,Lokhov2014,Koher2019}.
This method effectively predicts the probability of each node being in a given state at time $t$, particularly in tree-like networks, and is robust to various initial conditions.

First, we derive the exact equations that govern the DMP method.
Let $P^i_{S}(t)$, $P^i_{A}(t)$, and $P^i_{R}(t)$ denote the probabilities of node $i$ being in the susceptible ($S$), adopted ($A$), or recovered ($R$) state at time $t$, respectively.
These probabilities satisfy the following constraint:
\begin{equation}\label{eq:sum_to_one}
    P_A^i(t) + P_S^i(t) + P_R^i(t) = 1.
\end{equation}

The recovery of an adopted node $i$ occurs independently, regardless of its neighbors' states.
Therefore, the probability of node $i$ being in the recovered state at time $t$, denoted $P_R^i(t)$, is given by:
\begin{equation}\label{eq:eqs4}
	P_R^i(t) = P_R^i(t-1) + \mu P_A^i(t-1),
\end{equation}
where $\mu$ is the fixed recovery probability.
To find the overall fraction of recovered nodes in the network at time $t$, we sum $P_R^i(t)$ across all nodes:
\begin{eqnarray}\label{eq:eqs5}
	R(t) = \sum_{i=1}^N{P_R^i(t)}.
\end{eqnarray}
Inserting Eq.~(\ref{eq:eqs5}) into Eq.~(\ref{eq:eqs3}) yields:
\begin{eqnarray}\label{eq:eqs7}
	\beta'(t) = \min\left(1,\;\beta + \alpha\frac{\sum^{N}_{i=1}P^{i}_{R}(t-1)}{N}\right),
\end{eqnarray}
where $\beta'(t)$ increases with $R(t-1)$ until it reaches a maximum value of 1.

For a susceptible node $i$, the probability of remaining in the susceptible state up to time $t$, denoted $P_S^i(t)$, can be described as:
\begin{eqnarray}\label{eq:eqs8}
	P_S^i(t) =  P_S^i(0) \Phi_i(t),
\end{eqnarray}
where $\Phi_i(t)$ is the probability that node $i$ has not received successful transmissions from any adopted neighbors by time $t$.
The DMP approach assumes that the network is tree-like, so each neighbor of node $i$'s evolves independently.
However, if node $i$ is influenced by a neighbor node $z$ and switches to the adopted state ($A$), it may attempt to influence another neighbor node $z'$.
This interdependence introduces a challenge, as the transitions of nodes $z$ and $z'$ become correlated once $i$ adopts the contagion.

To address this issue, we assume that the focal node $i$ is in a cavity state and define $\theta^{z \rightarrow i}(t)$ as the probability that node $i$ has not been successfully influenced by node $z$ up to time $t$.
Consequently, $\Phi_i(t)$ can be factorized as $\prod_{z \in \partial i} \theta^{z \rightarrow i}(t)$, where $\partial i$ represents the set of neighbors of node $i$.
Substituting this expression into Eq.~(\ref{eq:eqs8}) gives:
\begin{eqnarray}\label{eq:eqs9}
	P_S^i(t) = P_S^i(0) \prod_{z\in \partial i} \theta^{z\rightarrow i}(t).
\end{eqnarray}

In message-passing, directionality is preserved, meaning that $\theta^{i\rightarrow z}(t) \neq \theta^{z\rightarrow i}(t)$ in undirected networks.
We treat each undirected edge as two directed edges pointing in opposite directions.
Initially, we set $\theta^{z \rightarrow i}(0) = 1$ for all edges in the network.
As time progresses, $\theta^{z \rightarrow i}(t-1)$ decreases as the contagion is transmitted from node $z$ to node $i$ with probability $\beta'(t)\phi^{z \rightarrow i}(t-1)$, where $\phi^{z \rightarrow i}(t-1)$ represents the probability that the adopted node $z$ has not yet passed the contagion to node $i$ by time $t-1$.
Accordingly, $\theta^{z \rightarrow i}(t)$ updates according to the rule:
\begin{eqnarray}\label{eq:eqs10}
	\theta^{z\rightarrow i}(t) = \theta^{z\rightarrow i}(t-1) - \beta'(t) \phi^{z\rightarrow i}(t-1).
\end{eqnarray}

To derive the expression for $\phi^{z \rightarrow i}(t)$, we first recognize that
$\phi^{z \rightarrow i}(t)$ decreases when node $z$ (currently in the adopted state $A$) recovers, or it transmits the contagion to node $i$, or both events occur simultaneously.
The probabilities for these events are $\mu$, $\beta'(t)$, and $\mu \beta'(t)$, respectively.
Additionally, $\phi^{z \rightarrow i}(t)$ can increase if node $z$, originally in the susceptible state $S$, switches to the adopted state $A$.
The rate of this change, $\Delta P^{z \rightarrow i}_S(t)$, can be expressed as the difference between the probabilities of node $z$ remaining susceptible at consecutive times, i.e.,
$P_S^{z \rightarrow i}(t-1) - P_S^{z \rightarrow i}(t)$.
Here, $P_S^{z \rightarrow i}(t)$ represents the probability of node $z$ remaining in state $S$ after interacting with the node $i$ in the cavity state.
Combining these elements, we have:
\begin{widetext}
\begin{eqnarray}\label{eq:eqs11}
		\phi^{z\rightarrow i}(t) & = & \phi^{z\rightarrow i}(t-1) - \beta'(t)\phi^{z\rightarrow i}(t-1) -\mu\phi^{z\rightarrow i}(t-1) + \mu\beta'(t)\phi^{z\rightarrow i}(t-1) \\
        & & + P_S^{z\rightarrow i}(t-1) - P_S^{z\rightarrow i}(t)\notag \\ \nonumber
		& = & (1-\beta'(t))(1-\mu)\phi^{z\rightarrow i}(t-1) + P_S^{z\rightarrow i}(t-1) - P_S^{z\rightarrow i}(t).
\end{eqnarray}
\end{widetext}

Next, we need to explicitly define $P_S^{z \rightarrow i}(t)$.
Since we are treating node $i$ as a cavity,
node $z$ will remain susceptible as long as it is not influenced by any of its neighbors other than $i$.
Using Eq.~(\ref{eq:eqs9}), the probability that node $z$ remains susceptible while $i$ is a cavity is given by:
\begin{eqnarray}\label{eq:eqs12}
	P_S^{z \rightarrow i}(t) = P_S^z(0) \prod_{j\in\partial z\setminus i} \theta^{j\rightarrow z}(t),
\end{eqnarray}
where $\partial z\setminus i$ represents the set of neighbors of $z$, excluding $i$.
At $t=0$, we set the initial condition as $P_S^{z\rightarrow i}(0)=1$ if $z$ is not the initial adopter, and $P_S^{z \rightarrow i}(0) = 0$ if $z$ is the seed node.
This can be compactly represented by $P_S^{z \rightarrow i}(0) = 1 - \delta_{q_z(0),A}$, where $\delta_{q_z(0),A}$ is the Kronecker delta function indicating the initial state of node $z$.

Using Eqs.~(\ref{eq:eqs7}), (\ref{eq:eqs10})-(\ref{eq:eqs12}), we can finally compute the trajectories of $\theta^{z \rightarrow i}(t)$, $\phi^{z \rightarrow i}(t)$, and $P_S^{z \rightarrow i}(t)$, starting with the following initial conditions:
\begin{eqnarray}\label{eq:eqs13}
\theta^{z\rightarrow i}(0) & = &  1, \\
\phi^{z\rightarrow i}(0) & = & P^z_A(0) = \delta_{q_z(0),A},\\
P^{z\rightarrow i}_S(0) & = & P^z_S(0) =  1-\delta_{q_z(0),A}.
\end{eqnarray}
Additionally, by incorporating the initial conditions $P_R^i(0)=0$ and $\beta'(0)=\beta$ into Eqs.~(\ref{eq:sum_to_one}), (\ref{eq:eqs4}), (\ref{eq:eqs7}), and (\ref{eq:eqs9})-(\ref{eq:eqs12}), we can compute the complete trajectories of $P^i_{S}(t)$, $P^i_{A}(t)$, $P^i_{R}(t)$, ultimately determining the order parameter $R(\infty)$.
The DMP approach used here has a computational complexity of $O(L)$, where $L$ denotes the number of edges in the network.

\section{Relationship between reinforcement threshold and localization strength}
To characterize the relationship between the localization strength
$\mathcal{L}$ and $\alpha_c$, we begin by calculating their values across
various network models and real-world networks (see Supplemental Material~\cite{SMA2}, Sec.~A.2) and visualizing them
We observe that $\alpha_c$ increases linearly with $\mathcal{L}$ on a logarithmic scale, suggesting a relationship of the form $\ln \alpha_c = \eta \ln \mathcal{L} + \delta$, or equivalently, $\alpha_c = \lambda \mathcal{L}^{\eta}$, where $\lambda = e^{\delta}$.
The parameters $\eta$ and $\delta$ are estimated from the data.
To avoid overfitting the real-world networks, we apply the least squares method to fit this relationship to a set of network models, yielding $\eta = 1.00$ and $\lambda = 2.63$.
Therefore, we have $\alpha_c = 2.63 \mathcal{L}$.
This relationship allows us to estimate $\alpha_c$ based on network topology.

\section{Relationship between critical point and reinforcement threshold}
With the analytical expressions of $\alpha_c$ and $\beta_c$, we can now explore their relationship.
In uncorrelated random networks, the components of the leading eigenvector in the non-backtracking matrix are determined by node degrees, as expressed in Eq.~(\ref{eq:eqs26})~\cite{Pastor2020},
\begin{eqnarray}\label{eq:eqs26}
    v_{j\rightarrow i} \sim k_j -1.
\end{eqnarray}
Using this approximation, we can represent the non-backtracking centrality for node $i$ as:
\begin{equation}\label{eq:eqs27}
    x_i = \sum_{j}A_{ij}v_{j \rightarrow i} \sim \sum_j{\frac{k_i k_j}{N \langle k \rangle}(k_j-1)} =\frac{\langle k^2 \rangle - \langle k \rangle}{\langle k \rangle} k_i,
\end{equation}
where we replace $A_{ij}$ by its expectation in annealed networks, $\hat{A}_{ij} = \frac{k_i k_j}{N \langle k \rangle}$.
The first moment of $x_i$ then becomes:
\begin{eqnarray}\label{eq:eqs28}
    \langle x_i \rangle = \frac{\sum_i x_i}{N} \sim \langle k^2 \rangle - \langle k \rangle.
\end{eqnarray}
Substituting Eqs.~(\ref{eq:eqs27}) and (\ref{eq:eqs28}) into the expression for localization strength, Eq.~(\ref{eq:localization_strength}), we obtain:
\begin{eqnarray}\label{eq:eqs29}
    \mathcal{L} = \frac{\sqrt{\langle k^2 \rangle - {\langle k \rangle}^2}}{{\langle k \rangle}^2}.
\end{eqnarray}
This result shows that $\mathcal{L}$ depends solely on the first- and second-order moments of $\langle k \rangle$.
Similarly, in annealed networks~\cite{Pastor2020}, $\beta_c$ can be written as:
\begin{eqnarray}\label{eq:eqs30}
    \beta_c = \frac{\langle k \rangle}{\langle k^2 \rangle - \langle k \rangle}.
\end{eqnarray}
Combining Eqs.~(\ref{eq:eqs29}) and (\ref{eq:eqs30}), we have:
\begin{eqnarray}\label{eq:eqs31}
   \beta_c = \frac{1}{\langle k\rangle^3 \mathcal{L}^2 +\langle k \rangle -1}.
\end{eqnarray}

Finally, by incorporating the relationship $\alpha_c = \lambda \mathcal{L}^\eta$, we can express $\beta_c$ in terms of $\alpha_c$:
\begin{eqnarray}\label{eq:eqs32}
   \beta_c = \frac{1}{\langle k\rangle^3 (\alpha_c/\lambda)^{2/\eta} +\langle k \rangle -1}.
\end{eqnarray}

\section*{Data availability}
All data supporting this study are available on Mendeley Data~(\url{https://data.mendeley.com/datasets/d848h7rcdg}) and are described in detail in the Supplemental Material~\cite{SMA1}.




\end{document}